\overfullrule=0pt
\input harvmac
\def\a{{\alpha}}
\def\ah{{\widehat\alpha}}

\def\l{{\lambda}}
\def\lh{{\widehat\lambda}}
\def\b{{\beta}}
\def\bh{{\widehat\beta}}
\def\g{{\gamma}}
\def\gh{{\widehat\gamma}}
\def\d{{\delta}}
\def\dh{{\widehat\delta}}

\def\N{{\nabla}}

\def\half{{1\over 2}}
\def\p{{\partial}}

\def\t{{\theta}}
\def\th{{\widehat\theta}}

\Title{ \vbox{\baselineskip12pt
\hbox{IFT-P.079/2000, UCCHEP/14-00}}}
{\vbox{\centerline
{Superstring Vertex Operators }
\centerline{in an $AdS_5\times S^5$ Background}}}
\smallskip
\centerline{Nathan Berkovits\foot{e-mail: nberkovi@ift.unesp.br}}
\smallskip
\centerline{\it 
Instituto de F\'{\i}sica Te\'orica, Universidade Estadual Paulista}
\centerline{\it
Rua Pamplona 145, 01405-900, S\~ao Paulo, SP, Brasil}
\bigskip
\centerline{Osvaldo Chand\'{\i}a\foot{e-mail: ochandia@maxwell.fis.puc.cl}}
\smallskip
\centerline{\it Facultad de F\'{\i}sica, Pontificia Universidad Cat\'olica
de Chile}
\centerline{\it Casilla 306, Santiago 22, Chile}

\bigskip

\noindent

A quantizable action has recently been proposed for the superstring
in an $AdS_5\times S^5$ background with Ramond-Ramond flux.
In this paper, we construct physical vertex operators corresponding
to on-shell fluctuations around the $AdS_5\times S^5$ background. The
structure of these $AdS_5\times S^5$ vertex operators closely
resembles the structure of the massless 
vertex operators in a flat background.
 
\Date{September 2000}

\newsec{Introduction}

A quantizable action has recently been proposed
for the superstring in an $AdS_5\times S^5$ background
with Ramond-Ramond flux \ref\berk {N. Berkovits, ``Super-Poincar\'e
Covariant Quantization of the Superstring,'' JHEP 04 (2000) 018,
hep-th/0001035.}. 
This action is closely related
to the $AdS_2\times S^2$ and $AdS_3\times S^3$ actions of 
\ref\adswitten
{N. Berkovits, C. Vafa and E. Witten, ``Conformal Field Theory
of $AdS$ Background with Ramond-Ramond Flux,'' JHEP 9903 (1999)
018, hep-th/9902098.}
\ref\bz
{N. Berkovits, M. Bershadsky, T. Hauer, S. Zhukov and B. Zwiebach,
``Superstring Theory on $AdS_2\times S^2$ as a Coset Supermanifold,''
Nucl. Phys. B567 (2000) 61, hep-th/9907200.}
\ref\six{N. Berkovits, ``Quantization of the
Type II Superstring in a Curved Six-Dimensional Background,''
Nucl. Phys. B565 (2000) 333, hep-th/9908041.}, 
and differs from the 
Green-Schwarz action of Metsaev and Tseytlin
\ref\metsaev{R. Metsaev and A. Tseytlin, ``Type
IIB Superstring Action in $AdS_5\times S^5$ Background,''
Nucl. Phys. B533 (1998) 109, hep-th/9805028.} in that
quantization is straightforward using a BRST charge constructed
out of the fermionic constraints and a pure spinor. 
Physical vertex operators can be described in a manifestly
spacetime-supersymmetric manner as states in the cohomology
of this BRST charge \berk
\ref\cohom{N. Berkovits, ``Cohomology in the Pure Spinor 
Formalism for the Superstring,'' hep-th/0006003.}. 
In an $AdS_3\times S^3$ background with
Ramond-Ramond flux, physical vertex operators corresponding
to on-shell fluctuations of the $AdS_3\times S^3$ background were
explicitly
constructed by Dolan and Witten in
\ref\dw{L. Dolan and E. Witten, ``Vertex Operators for
$AdS_3$ Background with Ramond-Ramond Flux,'' JHEP 9911 (1999) 003, 
hep-th/9910205 .}. In this paper, we shall generalize their construction
to the
$AdS_5\times S^5$ background.
The resulting $AdS_5\times S^5$
vertex operators will closely resemble the massless
Type IIB vertex operators in a flat ten-dimensional background.

In a flat background, the massless Type IIB vertex operators
in the pure spinor formalism for the superstring
are constructed using a bispinor 
superfield, $A_{\b \gh}(x,\t,\th)$, satisfying the equations of motion
and gauge invariances 
\eqn\efm{\g_{mnpqr}^{\a\b} D_\a A_{\b\gh} =
\g_{mnpqr}^{\ah\gh} D_\ah A_{\b\gh} =
0,}
$$
\d A_{\b\gh} = D_\b \widehat\Omega_\gh + \widehat
D_\gh \Omega_\b
{\rm ~~~with~~~}
\g_{mnpqr}^{\a\b} D_\a \Omega_\b =
\g_{mnpqr}^{\ah\gh} D_\ah \widehat\Omega_{\gh} = 0,$$
where $D_\a={\p\over{\p\t^\a}} + (\g^m\t)_\a \p_m$ and 
$ 
D_\ah={\p\over{\p\th^\ah}} + (\g^m\th)_\ah \p_m$ are the supersymmetric
derivatives of flat $N=2$ $D=10$ superspace and 
$\g^m_{\a\b}$ are the symmetric $16\times 16$ SO(9,1) Pauli
matrices.\foot{
Throughout this paper, we shall use the notation $\a$ and $\ah$
to represent the two real sixteen-component Majorana-Weyl
spinors of $N=2$ $D=10$ superspace. Although they transform as
the same SO(9,1) representation for the Type IIB superstring, it will
be convenient to use two separate labels for the two different spinors.}
As will be reviewed in section 2,
the components of $A_{\a\bh}$ which satisfy \efm\ describe the linearized
on-shell Type IIB supergravity fields.

As will be shown in section 3, the vertex operators for on-shell
fluctuations
around the $AdS_5\times S^5$ background can also be constructed using
a bispinor superfield,
$A_{\b \gh}(x,\t,\th)$, which now must satisfy the equations of motion
and gauge invariances
\eqn\eom{\g_{mnpqr}^{\a\b} \N_\a A_{\b\gh} =
\g_{mnpqr}^{\ah\gh} \N_\ah A_{\b\gh} =
0,}
$$
\d A_{\b\gh} = \N_\b \widehat\Omega_\gh + 
\N_\gh \Omega_\b {\rm ~~~with~~~}
\g_{mnpqr}^{\a\b} \N_\a \Omega_\b =
\g_{mnpqr}^{\ah\gh}  \N_\ah \widehat\Omega_{\gh} = 0,$$
where $\N_\a$ and 
$\N_\ah$ are the covariant supersymmetric derivatives in the
$AdS_5\times S^5$ superspace background. 
Note that in a generic
Type IIB supergravity background, the equations of motion
and gauge invariances of \eom\ would be inconsistent because 
$\{\N_\a, \N_\bh\}$ is non-vanishing. Although
this anti-commutator is also non-vanishing in the 
$AdS_5\times S^5$ background, the symmetrical structure of 
$\{\N_\a, \N_\bh\}$ in the $AdS_5\times S^5$ background 
is enough to make \eom\ consistent.

It is sometimes
convenient to choose a gauge in which the vertex operator is
a worldsheet primary field. In a flat background, the bispinor
superfield
$A_{\b\gh}$ satisfies in this gauge
\eqn\harmf{\p_n\p^n A_{\a\bh} =0,\quad \p^n A_{n\bh} = \p^n A_{\a n} =0,}
$$
\d A_{\b\gh} = \N_\b \widehat\Omega_\gh + 
\N_\gh \Omega_\b {\rm ~~~with~~~} \p_n\p^n \Omega_\b=
\p_n\p^n \widehat\Omega_\gh= \p^n \Omega_n =
\p^n \widehat\Omega_n= 0$$
where $A_{m\gh} = {1\over {16}}\g_m^{\a\b} D_\a A_{\b\gh}$,
$A_{\a m} = {1\over {16}}\g_m^{\bh\gh} D_\bh A_{\a\gh}$, 
$\Omega_m = {1\over {16}}\g_m^{\a\b} D_\a \Omega_\b$, 
$\widehat\Omega_m = {1\over {16}}\g_m^{\ah\bh} D_\ah \widehat\Omega_\bh$.
As will
be shown in section 4, there exists an analogous gauge choice
in the $AdS_5\times S^5$ background in which the bispinor superfield
$A_{\b\gh}$ satifies 
\eqn\harma{\N_B\N^B A_{\a\bh} =0,\quad \N^B A_{B\bh} = \N^B A_{\a B} =0,}
$$
\d A_{\b\gh} = \N_\b \widehat\Omega_\gh + 
\N_\gh \Omega_\b {\rm ~~~with~~~} \N_B\N^B \Omega_\b=
\N_B\N^B \widehat\Omega_\gh= \N^B \Omega_B =
\N^B \widehat\Omega_B= 0$$
where 
$B$ ranges over all tangent-space
indices of the super Lie-algebra of $PSU(2,2|4)$.
The equations of \harma\ are manifestly invariant under the AdS isometries 
and reduce in the flat limit to the equations of \harmf.

\newsec{Massless Vertex Operators in a Flat Background}

In this section, we review the construction of massless vertex
operators for the Type IIB superstring in a flat
background using the pure spinor formalism.
We will first review the manifestly super-Poincar\'e invariant worldsheet
action and will then discuss the physical massless vertex operators.

\subsec{Worldsheet action in a flat background}

In a flat background, 
the worldsheet action in conformal gauge 
in this formalism is
\eqn\flat{S=\int d^2 z [\half \p x^m \overline\p x_m +
p_\a\overline\p\t^\a
+ \widehat p_\ah \p\th^\ah 
] + S_{\l} + S_{\lh}.}
where $m=0$ to 9, $\a$ and $\ah=1$ to 16,
and $S_\l$ and $S_{\lh}$ are the worldsheet free-field actions\foot{ 
Although an explicit construction of $S_\l$ and $S_\lh$ 
requires breaking SO(9,1) to a subgroup, the OPE's of $\l^\a$
and $\lh^\ah$ with their Lorentz currents $N^{mn}$ and $\widehat N^{mn}$
are manifestly SO(9,1) covariant. This allows all vertex operator
and scattering amplitude computations
to be manifestly SO(9,1) super-Poincar\'e covariant\berk.}
for an independent
left and right-moving spinor variable, $\l^\a$ and $\lh^\ah$, satisfying 
the pure spinor conditions
\eqn\pure{ \l\g^m\l=0 {\rm ~~and ~~} \lh\g^m\lh=0
{\rm ~~for~~} m=0 {\rm ~to~}  9.}
Note that all unhatted worldsheet variables
(except for $x^m$) in \flat\ are left-moving and all hatted
worldsheet variables are right-moving.

The free-field action of \flat\ 
is manifestly invariant under 
$N=2$ $D=10$ spacetime-supersymmetry
and it is convenient
to define the following spacetime-supersymmetric combinations
of the worldsheet fields,
\eqn\defdtwo{d_\a=p_\a - (\Pi^m -{1\over 2}\t\g^m\p\t) (\g_m\t)_\a,
\quad \Pi^m = \p x^m + \t\g^m\p\t,}
$$ \widehat d_\ah=\widehat p_\ah -(\widehat\Pi^m -
{1\over 2}\th\g^m\overline\p\th) (\g_m\th)_\ah,
\quad \widehat\Pi^m = \overline\p x^m + \th\g^m\overline\p\th,$$
which satisfy the OPE's
\eqn\ope{d_\a(y) d_\b(z)\to -2 \g^m_{\a\b} (y-z)^{-1} \Pi_m,\quad
\widehat d_\ah(\overline y) \widehat d_\bh(\overline z)\to -2
\g^m_{\ah\bh} 
(\overline y-\overline z)^{-1} \widehat\Pi_m.}

\subsec{Physical massless vertex operators}

Physical states in the pure spinor formalism are defined as vertex
operators
of ghost-number $(1,1)$ in the cohomology of
the nilpotent BRST-like charges
\eqn\brst{Q = \oint dz \l^\a d_\a {\rm ~~~and~~~} \overline Q = \oint 
d\overline z
\lh^\ah \widehat d_\ah,} 
where $\l^\a$ 
and $\lh^\ah$ carry ghost-number $(1,0)$ and $(0,1)$ respectively.
The massless states are constructed from zero modes only and are
therefore
represented by the vertex operator
$U=\l^\b \lh^\gh A_{\b\gh}(x,\t,\th)$
where $A_{\b\gh}(x,\t,\th)$ is a bispinor
$N=2$ $D=10$ superfield which depends only on
the worldsheet zero modes of $x^m$, $\t^\a$ and $\th^\ah$.

The cohomology condition for a physical vertex operator $U$ is that
it satisfies the equations and gauge invariances\foot{
For the ordinary closed bosonic string with the standard definitions
of $Q$ and $\overline Q$,
these cohomology conditions reproduce the usual
physical spectrum for states with non-zero momentum. For zero momentum
states, there are additional subtleties associated with the $b_0-\overline
b_0$
condition which will be ignored in the discussion of this paper.}
\eqn\cohomd{
QU= \overline Q U=0,}
$$
\d U = Q\Lambda + \overline Q\widehat\Lambda {\rm ~~~with~~~ }
\overline Q\Lambda = Q\widehat\Lambda=0.$$
Applying these conditions to $U=\l^\b\lh^\gh A_{\b\gh}$ 
where $\Lambda=\lh^\ah \widehat\Omega_\ah$ and
$\widehat\Lambda=\l^\a \Omega_\a$,
and using that pure spinors satisfy
\eqn\using{\l^\a\l^\b = {1\over{1920}}(\l\g^{mnpqr}\l) \g_{mnpqr}^{\a\b}
{\rm ~~~ and ~~~}
\lh^\ah\lh^\bh = {1\over{1920}}(\lh\g^{mnpqr}\lh) \g_{mnpqr}^{\ah\bh},}
one finds that $A_{\b\gh}$ must satisfy the conditions 
\eqn\efms{\g_{mnpqr}^{\a\b} D_\a A_{\b\gh} =
\g_{mnpqr}^{\ah\gh} D_\ah A_{\b\gh} =
0,}
$$
\d A_{\b\gh} = D_\b \widehat\Omega_\gh + 
D_\gh \Omega_\b
{\rm ~~~with~~~}
\g_{mnpqr}^{\a\b} D_\a \Omega_\b =
\g_{mnpqr}^{\ah\gh}  D_\ah \widehat\Omega_{\gh} = 0,$$
where $D_\a={\p\over{\p\t^\a}} + (\g^m\t)_\a \p_m$ and 
$
D_\ah={\p\over{\p\th^\ah}} + (\g^m\th)_\ah \p_m$ are the supersymmetric
derivatives of flat $N=2$ $D=10$ superspace.
Note that $Q \Phi =\l^\a D_\a \Phi$ and
$\overline Q \Phi = \lh^\ah D_\ah \Phi$ 
for any superfield $\Phi(\l,\lh,x,\t,\th)$.

It will now be shown that the conditions of \efms\
correctly reproduce the Type IIB
supergravity spectrum. The easiest way to check this is
to use the fact that closed superstring vertex operators 
can be understood as the left-right product of open superstring
vertex operators. The massless open superstring vertex operator
is described by an 
$N=1$ $D=10$ spinor superfield $A_\b(x,\t)$ satisfying the conditions  
\eqn\sym{\g_{mnpqr}^{\a\b} D_\a A_\b(x,\t) =0,\quad\d A_\b(x,\t) = 
D_\b \Omega(x,\t).} 
The conditions of \sym\ imply the super-Maxwell
spectrum \ref\sieg{W. Siegel, ``Superfields in Higher-Dimensional
Spacetime,'' Phys. Lett. B80 (1979) 220\semi
E. Witten,
``Twistor-like Transform in Ten Dimensions,''
Nucl. Phys. B266 (1986) 245\semi
P. Howe, ``Pure Spinor Lines in Superspace and Ten-Dimensional
Supersymmetric Theories,'' Phys. Lett. B258 (1991) 141,
Addendum-ibid. B259 (1991) 511.}
since
there exists a gauge choice such that
\eqn\su{
A_\b(x,\t)= (\g^m\t)_\b a_m (x) + (\t\g^{mnp}\t) (\g_{mnp}\xi(x))_\b +...}
where
$a_m(x)$ and $\xi^\a(x)$ satisfy the super-Maxwell equations
$\p^m(\p_m a_n-\p_n a_m)= \p_m (\g^m\xi)_\a =0$ and where
$...$ contains only auxiliary fields.

Similarly, the equations of \efms\ imply that there exists a gauge choice
such
that 
\eqn\sugra{A_{\b\gh}(x,\t,\th) = (\g^m\t)_\b (\g^n\th)_\gh h_{mn}(x)
+ (\g^m\t)_\b (\th\g^{pqr}\th) (\g_{pqr}\widehat\psi_m(x))_\gh }
$$
+ (\t\g^{pqr}\t)(\g^m\th)_\gh (\g_{pqr}\psi_m(x))_\b
+ (\t\g_{mnp}\t)\g^{mnp}_{\b\a}(\th\g_{qrs}\th)\g^{qrs}_{\gh\dh} 
F^{\a\dh}(x)
+ ... $$
where $...$ contains only auxiliary fields and where
$h_{mn}(x)$, $\psi^\a_m(x)$, $\widehat\psi^\ah_m(x)$ and
$F^{\a\dh}(x)$ satisfy the equations 
\eqn\compeq{\p^m(\p_m h_{pn} -\p_p h_{mn})  =
\p^m(\p_m h_{pn} - \p_n h_{pm}) =0,}
$$\p^m (\p_m \psi^\a_n - \p_n\psi^\a_m) = 
\p^m (\p_m \widehat\psi^\ah_n - \p_n\widehat\psi^\ah_m) = 0,\quad
\p_n (\g^n \psi_m)_\a = 
\p_n (\g^n \widehat\psi_m)_\ah = 0,$$
$$\g^n_{\a\b} \p_n F^{\b\gh} = \g^n_{\gh\dh} \p_n F^{\a\dh}=0.$$
The equations of \compeq\ are those of linearized Type IIB supergravity
where 
$h_{mn}$ describes the dilaton, graviton and
anti-symmetric two-form, $\psi_m^\a$ and $\widehat\psi_m^\ah$ describe
the two gravitini and dilatini, 
and $F^{\a\dh}$ describes the Ramond-Ramond field
strengths.

\newsec{ Vertex Operator in $AdS_5\times S^5$ Background}

In this section, it will be shown how to generalize the
equations of \efms\ in an $AdS_5\times S^5$ background.
We shall first discuss the superstring action in an $AdS_5\times S^5$
background and will then discuss the physical vertex operators which
describe on-shell fluctuations of this background.

\subsec{Action in $AdS_5\times S^5$ background}

In
an $AdS_5\times S^5$ background with Ramond-Ramond flux, the worldsheet
action for the superstring in the pure spinor formalism is
\eqn\adsaction{S= 
\int d^2 z [ \half(\eta_{cd} J^c \overline J^d +\eta_{c'd'} J^{c'} 
\overline J^{d'})
+ \d_{\a\bh}
(3 J^\bh \overline J^\a - J^\a \overline J^\bh) }
$$+ N_{cd} \overline J^{[cd]} +N_{c'd'} \overline J^{[c'd']} +\widehat
N_{cd}
J^{[cd]}+\widehat N_{c'd'} J^{[c'd']} +\half( N_{cd}\widehat N^{cd}
-N_{c'd'}\widehat N^{c'd'})]  + S_\l + S_{\widehat\l}$$
where $J^A = (g^{-1} \p g)^A$ and $\overline J^A = (g^{-1} \overline\p
g)^A$ 
are
left-invariant currents constructed from the supergroup element
$g(x,\t,\widehat\t) \in PSU(2,2|4)$,
$[N^{cd},N^{c'd'}]$ and
$[\widehat N^{cd},\widehat N^{c'd'}]$ 
are the
$SO(4,1)\times SO(5)$ components of the Lorentz current 
for $\l^\a$ and
$\lh^\ah$, and $S_\l$ and $\widehat S_\lh$ are the same as in \flat. 
The indices $(c,[cd],c',[c'd'],\a,\ah)$ range over the 
tangent space indices of the super-Lie
algebra of $PSU(2,2|4)$ where $(c ,[cd])$ describe the SO(4,2) isometries
of $AdS_5$ with $c=0$ to 4,
and $(c',[c'd'])$ describe the SO(6) isometries of $S^5$ with $c'=5$ to 9.
It will be convenient to preserve the notation $\a$ and $\ah$ for the
two sixteen-component Majorana-Weyl spinors, but unlike the flat
case, one can now contract 
an $\a$ index with an $\ah$ index 
in an $SO(4,1)\times SO(5)$
invariant manner using 
the matrix $\d^{\a\ah}= \gamma_{01234}^{\a\ah}$ and
$\d_{\a\ah}= \gamma^{01234}_{\a\ah}$.
The non-vanishing structure constants $f_{AB}^C$ of the
$PSU(2,2|4)$ algebra are
\eqn\structure{
f_{\a\b}^{\underline{c}} =2 \g^{\underline{c}}_{\a\b},\quad
f_{\ah\bh}^{\underline{c}} =2 \g^{\underline{c}}_{\a\b},}
$$
f_{\a \bh}^{[ef]}= 
f_{\bh \a}^{[ef]}= 
(\g^{ef})_\a{}^\g \d_{\g\bh},\quad
f_{\a \bh}^{[e'f']}= 
f_{\bh \a}^{[e'f']}= -
(\g^{e'f'})_\a{}^\g \d_{\g\bh},$$
$$f_{\a \underline{c}}^\bh 
=-f_{\underline{c}\a}^\bh 
=\half (\g_{\underline c})_{\a\b}
\d^{\b\bh},\quad
f_{\ah \underline{c}}^\b = 
-f_{\underline{c}\ah}^\b = 
-\half
(\g_{\underline c})_{\ah\bh} \d^{\b\bh},$$
$$f_{c d}^{[ef]}= \d_c^{[e} \d_d^{f]},
\quad f_{c' d'}^{[e'f']}= -\d_{c'}^{[e'} \d_{d'}^{f']},$$
$$f_{[\underline{cd}][\underline{ef}]}^{[\underline{gh}]}=
\eta_{\underline{ce}}\d_{\underline{d}}^{[\underline{g}}
\d_{\underline{f}}^{\underline{h}]}
-\eta_{\underline{cf}}\d_{\underline{d}}^{[\underline{g}}
\d_{\underline{e}}^{\underline{h}]}
+\eta_{\underline{df}}\d_{\underline{c}}^{[\underline{g}}
\d_{\underline{e}}^{\underline{h}]}
-\eta_{\underline{de}}\d_{\underline{c}}^{[\underline{g}}
\d_{\underline{f}}^{\underline{h}]}$$
$$f_{[\underline{cd}] \underline{e}}^{\underline{f}} =
-f_{\underline{e} [\underline{cd}]}^{\underline{f}}= \eta_{\underline{e}
\underline{[c}} \d_{\underline d]}^{\underline{f}},\quad
f_{[\underline{cd}] \a}^{\b} =
-f_{\a [\underline{cd}]}^{\b}= \half(\g_{\underline{cd}})_\a{}^\b,\quad
f_{[\underline{cd}] \ah}^{\bh} =
-f_{\ah [\underline{cd}]}^{\bh}= \half(\g_{\underline{cd}})_\ah{}^\bh,$$
where $\underline{c}$ signifies either $c$ or $c'$ and 
$[\underline{cd}]$ signifies either $[cd]$ or $[c'd']$.

The action of \adsaction\
was derived in \berk\ by plugging the background values of the
$AdS_5\times S^5$ superfields into the sigma model action for the
superstring in a generic Type II background, and integrating out
the $d_\a$ and $\widehat d_\ah$ worldsheet fields as in
\adswitten\bz\six.\foot{The
$\half(N^{cd}\widehat N_{cd}-N_{c'd'}\widehat N^{c'd'})$ term 
in \adsaction\
comes from $R_{\underline{cd}\underline{ef}} N^{\underline{cd}}
\widehat N^{\underline{ef}}$ where 
$R_{\underline{cd}\underline{ef}}$ is the background Riemann tensor, and
was
mistakenly omitted in \berk. The need for such a term was first
pointed out by 
Bershadsky \ref\bersh{M. Bershadsky, private
communication.} based on one-loop conformal invariance arguments. Also,
we have changed our normalization of the second term in \adsaction\
so that our $PSU(2,2|4)$ structure constants agree with those of
\metsaev.}
This action is invariant
under $\d g= M g + g\Omega(x,\t,\th)$ where $M$ is
a global $PSU(2,2|4)$ transformation and $\Omega(x,\t,\th)$ is a local
$SO(4,1)\times SO(5)$ transformation which also rotates the pure
spinors as 
\eqn\rotat{\d\l^\a=\half\Omega^{[\underline c\underline d]}(\g_{\underline
c
\underline d}\l)^\a,\quad
\d\lh^\ah = 
\half\Omega^{[\underline c\underline d]}(\g_{\underline c
\underline d}\lh)^\ah,}
so $g(x,\t,\th)$ can be defined to take values
in the coset supergroup
${{PSU(2,2|4)}\over{SO(4,1)\times SO(5)}}$.

The first line of \adsaction\ is the ten-dimensional version
of the action proposed in \bz\ for $AdS_2\times S^2$.
As discussed in \bz, it is 
one-loop conformally invariant and
differs from the action of Metsaev and Tseytlin \metsaev\ in that
$\kappa$-symmetry is replaced by the condition that the
currents $J^\ah$ and $\overline J^\a$ are ``covariantly'' holomorphic
and anti-holomorphic. In other words, 
the equations of motion from the first
line imply that \bz 
\eqn\holom{\overline\p J^\ah = \half[\overline J^{[cd]} (\g_{cd} J)^\ah
+ \overline J^{[c'd']} (\g_{c'd'} J)^\ah]  {\rm ~~and~~}
\p \overline J^\a =\half [ J^{[cd]} (\g_{cd} \overline J)^\a
+ J^{[c'd']} (\g_{c'd'} \overline J)^\a].}
The second line of \adsaction\ is necessary so that the currents
$\d_{\a\ah}\l^\a J^\ah$ and 
$\d_{\a\ah}\lh^\ah J^\a$ satisfy
\eqn\hol{\overline\p(
\d_{\a\ah}\l^\a J^\ah) =0
{\rm ~~~and ~~~} \p (\d_{\a\ah}\lh^\ah J^\a)=0,}
which
will be used later for constructing the $AdS$ versions
of the BRST charges $Q$ and $\overline Q$.

To prove \hol, first note
that the equations of motion for $\l^\a$ and $\lh^\ah$
are 
\eqn \eoml
{\overline\p\l^\a= (\half\overline J^{[cd]} +{1\over 4}
 \widehat N^{cd}) (\g_{cd}\l)^\a
+(\half\overline J^{[c'd']} - {1\over 4}\widehat
N^{c'd'}) (\g_{c'd'}\l)^\a ,}
$$\p\lh^\ah= (\half J^{[cd]} + {1\over 4}
N^{cd}) (\g_{cd}\lh)^\ah
+(\half J^{[c'd']} - {1\over 4}
N^{c'd'}) (\g_{c'd'}\lh)^\ah.$$
If one includes the contribution from the second line, the equations
of motion of \holom\ for $J^\ah$ and $\overline J^\a$ get modified to
\eqn\holomt{\overline\p J^\ah =\half[
\overline J^{[cd]}(\g_{cd} J)^\ah
+ \overline J^{[c'd']}(\g_{c'd'} J)^\ah]}
$$ +{1\over 4}
[\widehat N^{cd} (\g_{cd} J)^\ah -\widehat N^{c'd'}(\g_{c'd'} J)^\ah
+ N^{cd} (\g_{cd}\overline J)^\ah - N^{c'd'}(\g_{c'd'}\overline J)^\ah],$$
$$
\p \overline J^\a =\half[ J^{[cd]}  (\g_{cd} \overline J)^\a
+ J^{[c'd']}(\g_{c'd'} \overline J)^\a ]$$
$$
+{1\over 4}
[N^{cd} (\g_{cd}\overline J)^\a - N^{c'd'}(\g_{c'd'}\overline J)^\a
+ \widehat N^{cd} (\g_{cd}J)^\a - \widehat N^{c'd'}(\g_{c'd'}J)^\a].$$
Putting \eoml\ and \holomt\ together, one finds
\eqn\resu{\overline\p(
\d_{\a\ah}\l^\a J^\ah )= -
{1\over 4}( N^{cd} (\g_{cd}\l)^\a 
-N^{c'd'}(\g_{c'd'}\l)^\a )\d_{\a\ah}\overline J^\ah ,}
$$\p(
\d_{\a\ah}\lh^\ah J^\a)=
-{1\over 4}( \widehat 
N^{cd} (\g_{cd}\lh)^\ah -\widehat N^{c'd'}(\g_{c'd'}\lh)^\ah )
\d_{\a\ah} J^\a .$$

It will now be argued that the right-hand sides of these two equations
vanish. As was explained in \berk, one can write
the Lorentz currents as
$N^{\underline c\underline d}=\half (\l\g^{
\underline c\underline d}w)$ and
$\widehat N^{\underline c\underline d}=\half (\lh\g^{
\underline c\underline d}\widehat w)$
where 
$w_\a$ and $\widehat w_\ah$ are
the conjugate
momenta to $\l^\a$ and $\lh^\ah$. 
And since $\l^\a\l^\b ={1\over{1920}}
(\l\g^{mnpqr}\l) \g_{mnpqr}^{\a\b}$ and
$\lh^\ah\l^\bh ={1\over{1920}}
(\l\g^{mnpqr}\l) \g_{mnpqr}^{\ah\bh}$,
the right-hand sides of \resu\
are proportional to 
\eqn\zeroo{\g_{cd}\g_{mnpqr}\g^{cd} -\g_{c'd'}\g_{mnpqr}\g^{c'd'}.}
But one can easily check that \zeroo\ vanishes for any choice of the
five-form direction $mnpqr$, implying that \hol\ is satisfied. 
The vanishing of \zeroo\ will also later be used to argue that the BRST
charges $Q$ and $\overline Q$ anti-commute.

\subsec{Physical vertex operators}

Since 
$\d_{\a\ah}\l^\a J^\ah$ and 
$\d_{\a\ah}\lh^\ah J^\a$ are holomorphic and anti-holomorphic, one
can construct the conserved charges
\eqn\brstads{Q= \oint dz \d_{\a\ah}\l^\a J^\ah, {\rm ~~~ and ~~~}
\overline Q= \oint d\overline z \d_{\a\ah}\lh^\ah \overline J^\a,}
which will be used to define physical vertex operators in 
the $AdS_5\times S^5$ background as was done for the
flat background in section 2. 
On-shell fluctuations 
around the $AdS_5\times S^5$ background are therefore described
by vertex operators of the form
$U=\l^a\lh^\ah A_{\a\ah}(x,\t,\th)$
satisfying \cohomd\ where $Q$ and $\overline Q$ are defined in \brstads.

To evaluate the implications of \cohomd, one has to know the OPE's
of $Q$ and $\overline Q$. Instead of trying to directly compute
these OPE's using the action of 
\adsaction, it will be simpler to deduce them from the
requirement that they preserve the $AdS$ isometries and that
they reduce correctly in the flat limit. As will be 
argued below, when acting on a ghost-number $(M,N)$ vertex operator 
$\Phi=
\l^{\a_1} ...\l^{\a_M} \lh^{\bh_1} ... \lh^{\bh_N}
A_{\a_1 ... \a_M \bh_1 ... \bh_N}(x,\t,\th),$
\eqn\ansatz{Q \Phi 
= \l^{\kappa}
\l^{\a_1} ...\l^{\a_M} \lh^{\bh_1} ... \lh^{\bh_N} \N_{\kappa}
A_{\a_1 ... \a_M \bh_1 ... \bh_N} {\rm ~~~ and ~~~}}
$$\overline Q \Phi
=\l^{\a_1} ...\l^{\a_M}\lh^{\widehat\kappa} \lh^{\bh_1} ... \lh^{\bh_N} 
\N_{\widehat\kappa}
A_{\a_1 ... \a_M \bh_1 ... \bh_N} $$
where 
$\N_\a=
E_\a^M(\p_M + \omega_M)$ and $\N_\ah= E_\ah^M(\p_M+\omega_M) $ are
the covariant supersymmetric derivatives in the $AdS_5\times S^5$
background, and $E_B^M$ and $\omega_M$ are the super-vierbein
and spin connection in the $AdS_5\times S^5$
background with $B$ ranging over the tangent-superspace
indices $(\underline{c},\a,\ah)$ 
and $M$
ranging over the curved superspace indices
$(m,\mu,\widehat\mu)$.
One can express $E_B^M$ and $\omega_M$ in terms of the coset
elements $g(x,\t,\th)$ by defining 
$E_B^M =(E_M^B)^{-1}$ 
and $\omega_M^{[\underline{cd}]} = E_M^{[\underline{cd}]}$ where 
$(g^{-1}dg)^B = E_M^B dX^M$ for 
$X^M=(x^m,\t^\mu,\th^{\widehat\mu})$. 

To justify \ansatz, one can check that it is the unique definition
which preserves all $AdS$ isometries and reduces
to $Q\Phi = \l^\a D_\a\Phi$ and 
$\overline Q\Phi = \lh^\ah  D_\ah\Phi$ in the flat limit.
Furthermore,
\ansatz\ is consistent with the nilpotency conditions
$Q^2 = \overline Q^2 = \{Q, \overline Q\}=0$.
The conditions $Q^2\Phi = \overline Q^2\Phi=0$ follow
from the fact that 
$\g_{mnpqr}^{\a\b}\{\N_\a,\N_\b\}=
\g_{mnpqr}^{\ah\bh}
\{\N_\ah,\N_\bh\}=0$. 
Although 
$\{\N_\a,\N_\bh\}$
is non-vanishing, its symmetrical structure allows
$\{Q, \overline Q\}\Phi $ to vanish since
\eqn\nilp{ \{Q, \overline Q\} \Phi 
= \l^{\kappa} \lh^{\widehat\tau}
\l^{\a_1} ...\l^{\a_M} \lh^{\bh_1} ... \lh^{\bh_N} 
\{\N_{\kappa},\N_{\widehat\tau}\}
A_{\a_1 ... \a_M \bh_1 ... \bh_N} }
\eqn\nilq{= \l^{\kappa} \lh^{\widehat\tau}
\l^{\a_1} ...\l^{\a_M} \lh^{\bh_1} ... \lh^{\bh_N} 
((\g^{cd})_{\kappa}{}^{\sigma} \d_{\sigma\widehat\tau} \N_{[cd]}-
(\g^{c'd'})_\kappa{}^\sigma \d_{\sigma\widehat\tau} \N_{[c'd']})
A_{\a_1 ... \a_M \bh_1 ... \bh_N}, }
where $\N_{[\underline{cd}]}$ acts as a Lorentz rotation in the
$[\underline{cd}]$ direction on the
$M+N$ spinor indices of 
$A_{\a_1 ... \a_M \bh_1 ... \bh_N}$, i.e.
\eqn\iea{\N_{[\underline{cd}]}A_{\a_1 ... \a_M \bh_1 ... \bh_N}=
\half[(\g_{\underline{cd}})_{\a_1}{}^\g
A_{\g\a_2 ...  \bh_N}
+(\g_{\underline{cd}})_{\a_2}{}^\g
A_{\a_1 \g \a_3 ... \bh_N}
+ ... + (\g_{\underline{cd}})_{\bh_N}{}^\gh
A_{\a_1 ... \bh_{N-1}\gh} ].} 
But since all indices
of $A_{\a_1 ... \a_M \bh_1 ... \bh_N}$
are contracted with either $\l^\a$ or $\lh^\ah$, 
one can use 
\using\
to argue that 
all terms in \nilq\ are proportional to \zeroo\ which identically
vanishes.
So $\{Q, \overline Q\}\Phi=0 $ as desired.

Using \ansatz, 
\cohomd\ implies that the bispinor superfield $A_{\a\bh}(x,\t,\th)$
in the physical vertex operator $U=\l^\a\lh^\bh A_{\a\bh}$
must satisfy the equations of motion and gauge invariances
\eqn\eomtwo{\g_{mnpqr}^{\a\b} \N_\a A_{\b\gh} =
\g_{mnpqr}^{\ah\gh} \N_\ah A_{\b\gh} =
0,}
$$
\d A_{\b\gh} = \N_\b \widehat\Omega_\gh + 
\N_\gh \Omega_\b {\rm ~~~with~~~}
\g_{mnpqr}^{\a\b} \N_\a \Omega_\b =
\g_{mnpqr}^{\ah\gh} \N_\ah \widehat\Omega_{\gh} = 0.$$
Although one could do a component analysis to check
that \eomtwo\ correctly describes
the on-shell fluctuations around the $AdS_5\times S^5$ background,
this is guaranteed to work since \eomtwo\
are the unique equations of motion and gauge invariances
which are invariant under the $AdS$ isometries and which reduce to the
massless Type IIB supergravity equations of \efms\ in the flat limit.

\newsec{Primary Vertex Operators}

It is sometimes convenient to choose a gauge for physical vertex 
operators such that they are dimension zero 
worldsheet primary fields, i.e. they
have no double poles with the stress tensor. For the ordinary closed
bosonic string, this is the gauge where $b_0 U = \overline b_0 U=0$.
In the pure spinor formalism, there is no natural candidate for the
$b$ ghost, so one needs to analyze the stress-tensor to find this
gauge-fixing condition.
This will be done first in a flat background and then in the
$AdS_5\times S^5$ background.

\subsec{Primary vertex operators in a flat background}

In a flat background, the left and right-moving stress tensors
are
\eqn\st{T = \half\p x^m \p x_m + p_\a \p\t^\a + T_\l {\rm~~and~~}
\overline T = \half\overline\p x^m \overline\p x_m + 
\widehat p_\ah \widehat\p\t^\ah + \overline T_\lh,}
where $T_\l$ and $\overline T_\lh$ are the $c=22$ stress-tensors 
constructed from the
pure spinor variables $\l^\a$ and $\lh^\ah$ \berk. When acting on the
massless
vertex operator $U=\l^\a\lh^\bh
A_{\a\bh}(x,\t,\th)$, 
the condition of no double poles with $T$ or $\overline T$
implies that $\p_m \p^m A_{\a\bh}=0$. Furthermore, the on-shell conditions
of \efm\ imply that 
$\p^m(\p_m A_{n \bh} -\p_n A_{m \bh})=
\p^m(\p_m A_{\a n} -\p_n A_{\a m})= 0$
where $A_{m\gh} = {1\over {16}}\g_m^{\a\b} D_\a A_{\b\gh}$ and 
$A_{\a m} = {1\over {16}}\g_m^{\bh\gh} D_\bh A_{\a\gh}$.
So the gauge-fixed equations for $A_{\a\bh}$ are
\eqn\gf{
\p^m\p_m A_{\a \bh} = 0,\quad
\p^m A_{m \bh}=
\p^m A_{\a m}= 0.}

Under the gauge transformation
$\d A_{\b\gh} = D_\b \widehat\Omega_\gh + \widehat
D_\gh \Omega_\b$, 
$\d A_{m \bh} = \p_m \widehat\Omega_\bh -  D_\bh \Omega_m$ and
$\d A_{\a m} = \p_m \Omega_\a -  D_\a \widehat \Omega_m$
where $\Omega_m = 
{1\over {16}}\g_m^{\a\b} D_\a \Omega_\b$ and 
$\widehat\Omega_m = 
{1\over {16}}\g_m^{\ah\bh} D_\ah \Omega_\bh$.
So the residual gauge transformations
which leave \gf\ invariant are
\eqn\residf{\d A_{\b\gh} = D_\b \widehat\Omega_\gh + 
D_\gh \Omega_\b
{\rm ~~~with~~~}
\p_m\p^m \Omega_\b =
\p_m\p^m \widehat\Omega_\bh =
\p^m \Omega_m = 
\p^m\widehat\Omega_m = 0.}
The conditions of \gf\ and \residf\ will now be generalized in
an $AdS_5\times S^5$ background.

\subsec{Primary vertex operators in an $AdS_5\times S^5$ background}

In the $AdS_5\times S^5$ background, the left and right-moving stress
tensors 
associated with the action of \adsaction\ are
\eqn\adsst{T = 
\half(\eta_{cd} J^c J^d +\eta_{c'd'} J^{c'} J^{d'})
- 4\d_{\a\bh} J^\a J^\bh 
+ N_{cd} J^{[cd]} +N_{c'd'} J^{[c'd']} + T_\l,}
$$\overline T = 
\half(\eta_{cd} \overline J^c \overline J^d +\eta_{c'd'} 
\overline J^{c'} \overline J^{d'})
- 4\d_{\a\bh} \overline J^\a \overline J^\bh 
+\widehat  N_{cd} \overline J^{[cd]} +
\widehat N_{c'd'} \overline J^{[c'd']} + \overline T_\lh $$
where $T_\l$ and $\overline T_\lh$ are defined as in \st.
As was done earlier with $Q$ and $\overline Q$, instead of
directly computing the OPE's of $T$ and $\overline T$, it will be simpler
to deduce them from the requirements that they preserve the $AdS$
isometries and reduce correctly in the flat limit.

When acting on the physical 
vertex operator $U=\l^\a\lh^\bh
A_{\a\bh}(x,\t,\th)$, 
the condition of no double poles with $T$ or $\overline T$
implies that  
\eqn\imppl{\N_B\N^B  
A_{\a\bh}(x,\t,\th)=0}
where
$\N_B\N^B = \eta^{BC}\N_B\N_C $, $B=
(\underline{c},[\underline{cd}],\a,\ah)$ ranges over all tangent space
indices of $PSU(2,2|4)$,
$\N_B = E_B^M (\p_M + \omega_M)$ are the covariant derivatives
in the $AdS_5\times S^5$ background when $B=(c,c',\a,\ah)$,
and
$\N_{[\underline{cd}]}$
acts as a Lorentz rotation in the
$\underline{cd}$ direction on all tangent
space indices. The only non-zero components of $\eta^{BC}$
are 
\eqn\defeta{\eta^{\a\bh} =-\eta^{\bh\a}= -{1\over 4}\d^{\a\bh}, \quad
\eta^{[cd][ef]} = -{1\over 4}\eta^{c[e}\eta^{f]d},\quad
\eta^{[c'd'][e'f']} = {1\over 4}\eta^{c'[e'}\eta^{f']d'},}
and $\eta^{\underline{c}\underline{d}}$ is the $SO(1,9)$ Minkowski
metric.
Note that $[\N_A,\N_B\} = f_{AB}^C\N_C $ where
$f_{AB}^C$ are the $PSU(2,2|4)$ structure constants
defined in \structure, and $[\N_B\N^B, \N_C]=0$ for all $C$.
In the flat limit,
$\N_B\N^B$ reduces to $\p_n\p^n$ since $\eta^{\underline{cd}}$
is the only surviving component of $\eta^{BC}$.

To find the $AdS_5\times S^5$ analog of 
\gf, it will be convenient to define $A_{B\gh}$ by
\eqn\defin{A_{\underline{c}\gh} = {1\over {16}}\g_{\underline{c}}^{\a\b}
\N_\a A_{\b\gh},\quad
A_{\ah\gh} = {1\over {5}}\d^{\b\dh}[
\g^c_{\ah\dh} (\N_c A_{\b\gh} -\N_\b A_{c\gh})
+\g^{c'}_{\ah\dh} (\N_{c'} A_{\b\gh} -\N_\b A_{c'\gh})], }
$$
A_{[cd]\gh} = \N_{c} A_{d\gh}
-\N_{d} A_{c\gh}, \quad
A_{[c'd']\gh} = -\N_{c'} A_{d'\gh} 
+\N_{d'} A_{c'\gh}.$$
Under the gauge transformation $\d A_{\a\bh}=\N_\a\widehat\Omega_\bh
+\N_\bh\Omega_\a$, one can check that $A_{B \bh}$ transforms as
\eqn\gtr{\d A_{B\bh}= \N_B \widehat\Omega_\bh -(-1)^{(B)} \N_\bh\Omega_B
+(-1)^{(E)} f_{C \bh}^D \eta^{CE} \eta_{BD} \Omega_E}
where $(B)=0$ if $B$ is a bosonic index, $(B)=1$ if $B$ is 
a fermionic index,
and
\eqn\ome{\Omega_{\underline c} = 
{1\over {16}}\g_{\underline{c}}^{\a\b}
\N_\a \Omega_{\b},\quad
\Omega_{\gh} = 
{1\over {5}}\d^{\b\dh}[
\g^c_{\ah\dh} (\N_c \Omega_{\b} -\N_\b \Omega_{c})
+\g^{c'}_{\ah\dh} (\N_{c'} \Omega_{\b} -\N_\b \Omega_{c'})],
\quad\Omega_{[\underline{cd}]}=0.}
One can similarly define $A_{\a B}$ 
as
\eqn\defint{A_{\g\underline{c}} = {1\over {16}}
\g_{\underline{c}}^{\ah\bh}
\N_\ah A_{\g\bh},\quad
A_{\g\a} = {1\over {5}}\d^{\d\bh}[
\g^c_{\a\d} (\N_c A_{\g\bh} -\N_\bh A_{\g c})
+\g^{c'}_{\a\d} (\N_{c'} A_{\g\bh} -\N_\bh A_{\g c'})], }
$$
A_{\g [cd]} = \N_{c} A_{\g d}
-\N_{d} A_{\g c},\quad
A_{[c'd']\gh} = -\N_{c'} A_{\g d'} 
+\N_{d'} A_{\g c'},$$
which transforms as
\eqn\gtrt{\d A_{\a B}= \N_B \Omega_\a -(-1)^{(B)} \N_\a\widehat\Omega_B
+(-1)^{(E)} f_{C\a }^D \eta^{CE} \eta_{BD}\widehat \Omega_E {\rm
~~~where}}
\eqn\omet{\widehat\Omega_{\underline c} = 
{1\over {16}}\g_{\underline{c}}^{\ah\bh}
\N_\ah \widehat\Omega_{\bh},\quad
\widehat\Omega_{\a} = 
{1\over {5}}\d^{\d\bh}[
\g^c_{\a\d} (\N_c \widehat\Omega_{\bh} -\N_\bh \widehat\Omega_{c})
+\g^{c'}_{\a\d} (\N_{c'} \widehat\Omega_{\bh} -\N_\bh
\widehat\Omega_{c'})],
\quad\widehat\Omega_{[\underline{cd}]}=0.}

Then the unique conditions on $A_{\a\bh}$ which preserve
the $AdS$ isometries and which reduce to \gf\ in the flat limit are
\eqn\gfads{
\N^B\N_B A_{\a \bh} = 0,\quad
\N^B A_{B \bh}=
\N^B A_{\a B}= 0.}
Furthermore, using the gauge transformations of \gtr\ and \gtrt\
one can check that these conditions are invariant under the residual
gauge transformations
\eqn\residfads{\d A_{\b\gh} = \N_\b \widehat\Omega_\gh + 
\N_\gh \Omega_\b
{\rm ~~~with~~~}
\N_B\N^B \Omega_\a =
\N_B\N^B \widehat\Omega_\ah =
\N^B \Omega_B = 
\N^B\widehat\Omega_B = 0,}
which reduce to \residf\ in the flat limit.
So the conditions of \gfads\ and \residfads\ for the primary vertex
operator
describing on-shell fluctuations of the $AdS_5\times S^5$
background closely resemble the conditions of \gf\ and \residf\
for the primary
vertex operator describing the massless Type IIB supergravity fields in
a flat background.

\newsec{Concluding Remarks}

In this paper we have shown that the structure of physical
vertex operators for on-shell fluctuations around the
$AdS_5\times S^5$ background closely resembles the structure of 
massless vertex operators in a flat background. This is true
both for the gauge-invariant form of the vertex operators and
for the gauge-fixed form of the primary vertex operators.

The next step is to compute
tree-level scattering amplitudes involving the $AdS_5\times S^5$
vertex operators. Since massless tree amplitudes are
straightforward to compute in a flat background using the pure
spinor formalism \berk\ref\val{N. Berkovits and B.C. Vallilo,
``Consistency of Super-Poincar\'e Covariant Superstring
Tree Amplitudes,'' JHEP 07 (2000) 015, hep-th/0004171.}, 
it is quite encouraging that 
the physical vertex
operators corresponding to on-shell fluctuations 
around the $AdS_5\times S^5$ background are described
by superfields which closely resemble those in
a flat background. Of course, to compute superstring scattering
amplitudes,
one also needs to know the OPE's of the worldsheet fields.
In a flat background, they are free-field OPE's, however, 
in an $AdS_5\times S^5$ background, the OPE's are
complicated because of non-holomorphicity of the currents. Nevertheless,
it might be possible to deduce their form by
requiring that they are invariant under the AdS isometries and that
they reduce to the appropriate free-field OPE's in the flat limit.

{\bf Acknowledgements:} 
NB would like to thank Michael Bershadsky, Louise Dolan, Juan
Maldacena, Hirosi Ooguri,
Warren Siegel, Cumrun Vafa and Edward Witten for useful
discussions, and CNPq grant 300256/94-9 and FAPESP grant
99/12763-0 for partial financial support.
OC
would like to thank FONDECYT grant 3000026 and FAPESP grant
98/02380-3 for financial support. 
This research was partially conducted during the period that NB
was employed by the Clay Mathematics Institute as a CMI Prize Fellow. 

\listrefs

\end